\documentclass[runningheads]{svmult}

\usepackage{makeidx}  
\usepackage{graphicx} 
\usepackage{subeqnar} 
\usepackage{multicol} 
\usepackage{physprbb} 
\makeindex       

\def\Mdot{\hbox{$\dot {\rm M}$}}

\def\Lsun{\hbox{L$_\odot$}}

\def\Minit{\hbox{M$_{\rm init}$}}
\def\Msun{\hbox{M$_\odot$}}

\def\Msunyr{\hbox{M$_\odot\,$yr$^{-1}$}}

\def\Vinf{\hbox{$v_\infty$}}
\def\kms{\hbox{km$\,$s$^{-1}$}}

\def\arcsec{\hbox{$^{\prime\prime}$}}
\def\farcs{\hbox{$.\!\!^{\prime\prime}$}}

\def\simgr{\mathrel{\hbox{\rlap{\hbox{\lower4pt\hbox{$\sim$}}}\hbox{$>$}}}}
\def\simls{\mathrel{\hbox{\rlap{\hbox{\lower4pt\hbox{$\sim$}}}\hbox{$<$}}}}

\begin{document}
\title*{The Initial Mass Function in the Galactic Center}
\toctitle{The Initial Mass Function in the Galactic Center}
%
%
\titlerunning{The Initial Mass Function in the Galactic Center}
%
\author{Donald F. Figer}
\authorrunning{Donald F. Figer}
%
%
\institute{Space Telescope Science Institute, Baltimore MD 21218; \\ Johns Hopkins University, Baltimore, MD 21218}

\maketitle       

\begin{abstract}
The Galactic Center contains the most massive young clusters in the
Galaxy and serves as the closest example of a massive starburst
region. Our recent results suggest that the Galactic Center
environment produces massive clusters with relatively flat
initial mass functions, as might be expected on theoretical grounds.
I will discuss these recent results,
along with evidence for star formation in the immediate
vicinity of the super massive black hole at the Galactic
Center. The results of this work might be useful in extrapolating
to other galactic centers with similar conditions, as well as
other starburst regions. 
\end{abstract}

\section{Introduction}
The Galactic Center region spans the central few hundred parsecs of the
Galaxy and represents about 0.04\% of the volume in the Galactic disk. 
Although small in size, the region contains 10\% of the molecular material and young stars
in the Galaxy, as evidenced by molecular line and radio maps (\cite{mor96}).  
The molecular clouds in the center tend to have high densities that are greater than 
10$^4$~cm$^{-3}$; more rarefied clouds are destroyed by the strong tidal
field produced by the large stellar mass peaking in the central parsec. In addition,
the infrared energy distributions produced by the clouds indicate very high
temperatures, T$_{\rm cloud}$~=~70~K; and molecular line observations indicate
extraordinary turbulent velocities (FWHM) of about 25~km~s$^{-1}$, even on scales 
as small as 0.3~pc. In accordance with approximate pressure balance, the
magnetic field strengths appear to also be high, on the order of a milliGauss.
\cite{mor96} provides an excellent review of the GC environment. 

Despite this tumultuous environment, the
GC contains stars which produce L~=~10$^9$~\Lsun, and 
L$_{\rm Ly-cont}$~=~(1$-$3)(10$^{52}$)~s$^{-1}$, values dominated by older
stars in the case of the former and younger stars in the case of the latter. 
Given this fertile breeding ground for young stars, we embarked on a program to
identify all the sites of recent star formation in the GC and measure the
initial mass function for particular sites, where possible (\cite{fig95a}; 
\cite{fig95b}; \cite{fig96}; \cite{fig98}; \cite{fig99a}; \cite{fig99b}). This
paper gives a summary of the results obtained so far, with a particular
emphasis on the most recent result, a measurement of the initial mass function (IMF)
in the Galactic Center. 

\section{Star Formation in the Galactic Center}

The environment described in the introduction would appear to be somewhat
hostile for star formation. Indeed, \cite{mor93} notes that the Jean's mass in the
GC is extraordinarily large ($\sim$10$^5$~\Msun). Such high Jean's
masses are at odds with the stellar observations, but the turbulent velocities are measured on size scales 
$\sim$0.3~pc, a size that corresponds to 30~\Msun\ (assuming n$\sim$10$^4~cm^{-3}$).
One way to avoid this inconsistency between the high Jean's mass and
the existence of stars with more normal masses is to invoke highly compressive 
events, i.e. cloud-cloud collisions. 

\subsection{Major Sites of Star Formation}
There are three young massive clusters in the GC: the Central cluster 
(\cite{bec68}, \cite{for87}, \cite{kra91}, \cite{all94}), \cite{lib95}, \cite{tam96}, \cite{gen00}),
the Arches cluster (\cite{nag95}; \cite{cot96};
\cite{ser98}; \cite{fig99a}), and the Quintuplet cluster (\cite{oku90};
\cite{nag90}; \cite{gla90}; \cite{fig99b}). They are 
nearly identical to each other in stellar content, except for age, the Arches
being about half the age (2~Myr) of the other two. 

The Central cluster refers to the collection of young ($\tau_{\rm age}<10~Myr$)
stars in the central parsec. In the previous two decades, over 30 evolved massive 
stars having \Minit$>$20~\Msun\ have been identified in this cluster, including
9 WR stars, 20 stars with Ofpe/WN9-like {\it K}-band spectra {\cite{naj97}), 
several red supergiants {\cite{ram00}), and
many luminous mid-infrared sources (\cite{gen96}). The detailed spectroscopic analyses 
by \cite{naj94} \cite{naj97} found that the brightest
HeI emission-line stars in the GC have 
extremely strong stellar winds ($\Mdot\sim$ 5 to 80 $\times 10^{-5}\, \Msunyr$) and
relatively small outflow velocities (\Vinf~$\sim$~300 to 1,000~\kms). These
values, taken together with the inferred stellar temperatures (17,000\,K to 30,000\,K)
and luminosities (1 to $30~\times~10^{5}$~\Lsun) suggest that the young cluster is about
4-6~Myr old and provides the majority of the ionizing flux (10$^{50.5}~s^{-1}$) and 
heating (10$^{7.5}$~\Lsun) which is evident in the central parsec (\cite{kra95}).

The Arches contains about 160 O-stars ($\Minit>20~\Msun$), more than any other cluster
in the Galaxy, and has a total mass of ~10$^4~\Msun$ (\cite{fig99a}). \cite{cot96}
and \cite{nag95} first recognized the cluster for its dozen or so massive emission-line
stars. Futher imaging by \cite{ser98} revealed that the cluster contains a large
number of O-stars and is responsible for heating and ionizing the surface of a 
nearby molecular cloud. A determination of the cluster age (2~Myr) and IMF 
was subsequently made by \cite{fig99a} using HST/NICMOS observations. 

The Quintuplet cluster has about 2(10$^4~\Msun$) in stars (\cite{fig99a}),
but its massive stars are more evolved than those in the Arches (\cite{fig99b}; 
\cite{fig99c}), consistent with 4~Myr age of the former. Indeed, 
the cluster contains more Wolf-Rayet stars than any cluster in the Galaxy, and
over a dozen stars in earlier stages of evolution, i.e., LBV, Ofpe/WN9, and OB supergiants. 
Further, it contains two ``Luminous Blue Variables \cite{geb00},'' including the ``Pistol Star.''
The ionizing flux from the cluster is 10$^{50.9}$ photons s$^{-1}$, or roughly 
what is required to ionize the nearby ``Sickle'' HII region (G0.18$-$0.04). 
The total luminosity from the massive cluster stars is $\approx$ 10$^{7.5}$ \Lsun,
enough to account for the heating of the nearby molecular cloud, M0.20$-$0.033. 
The cluster is unique in the Galaxy in stellar content and 
age, except, perhaps, for the young cluster in the central parsec of the Galaxy. 

\subsection{Star Formation Near the Black Hole}
\cite{fig00} presented 
2.0$-$2.4 $\mu$m spectroscopy of the central 0.1 square arcseconds of the 
Galaxy, showing that the brightest stars in this region
are hotter than typical red giants (also see \cite{gen97} and \cite{eck99}). Taken together with photometry and colors,
they concluded that these objects are likely OB main sequence stars. 
It is unlikely that the stars 
are late-type giants stripped of their outer envelopes because such sources would be 
much fainter than those observed. In addition, \cite{rie00}, \cite{bai99}, and \cite{ale99}
note that these objects are not likely to be remnants of stellar mergers, unless
the merger rate is elevated with respect to what is currently estimated. 
Given their location very near to the black hole, and their apparent relative
youth, it is unlikely that they have drifted inward via dynamical friction.
If we assume such a young age, and high stellar mass per star, then we are faced
with the fact that large amounts of mass (20~\Msun) have been compressed to very
high densities ($>$4(10$^{11}$)~cm$^{-3}$) very near to the black hole (several
thousand AU). 

\section{The IMF in the Galactic Center}

The IMF describes the relative number of stars produced in a star forming event
as a function of initial stellar mass. It is often expressed as a single power law
over mass ranges above 1 \Msun, with a form d(Log N)/d(Log \Minit) 
= ${\rm \Gamma}$ = $\alpha$ + 1;
${\rm \Gamma}$ = $-$1.35 for the Salpeter case \cite{sal55}. The IMF for most young clusters
can be described by a power law with Salpeter index, although significant variations
are evident ($-$0.7~$>$~${\rm \Gamma}$~$>$~$-$2.1) \cite{sca98}. Whether these variations
are statistically significant, and whether environmental factors affect the IMF, 
remains to be proven. 

{\it Hubble Space Telescope} ({\it HST}) Near-infrared Camera 
and Multi-object Spectrometer (NICMOS)
observations of the Arches and Quintuplet clusters were obtained \cite{fig99a} and
used to identify main sequence stars in the Galactic Center 
with initial masses well below 10~\Msun, leading to 
the first determination of the initial mass function (IMF) for any 
population in the Galactic Center. 
The mass function derived for the Arches cluster is shown in Figures 1. 
The central region (r~$<$~3\arcsec) is excluded from the analysis because our statistics are
already less than 50\% complete there for \Minit~$<$~35~\Msun; however, note the dotted-line
histogram which does include the inner region for the most massive stars. For the annulus with 
3\arcsec~$<$~r~$<$~9\arcsec, we find a slope
which is significantly greater than $-$1.0, and so is one of the flattest mass functions 
ever observed for \Minit~$>$~10~\Msun. 
If all of the data are forced to fit a single line, the slope is near $-$0.6,
but we note that interesting structure may be present in the mass function in the form of near-zero-slope plateaus
for 15~\Msun~$<$~\Minit~$<$~50~\Msun, and for \Minit~$>$~50, i.e.,  
the intrinsic mass function may be relatively flat at higher masses.   
In comparison, the average IMF slope for 30 clusters in the 
Milky Way and LMC is $\approx$ $-$1.4
for log(\Minit)~$>$~0.7 , although a few clusters have ${\rm \Gamma}$ $\approx$ $-$0.7\cite{sca98}.
Some of these clusters also suggest a flattening of the IMF at higher masses, although the
actual slopes in these comparison clusters are in general much more biased toward lower masses.
Finally, we find that there are $\simgr$10 stars with \Minit~$>$~120~\Msun. 

\begin{figure}[1]
\begin{center}
\includegraphics[width=.7\textwidth]{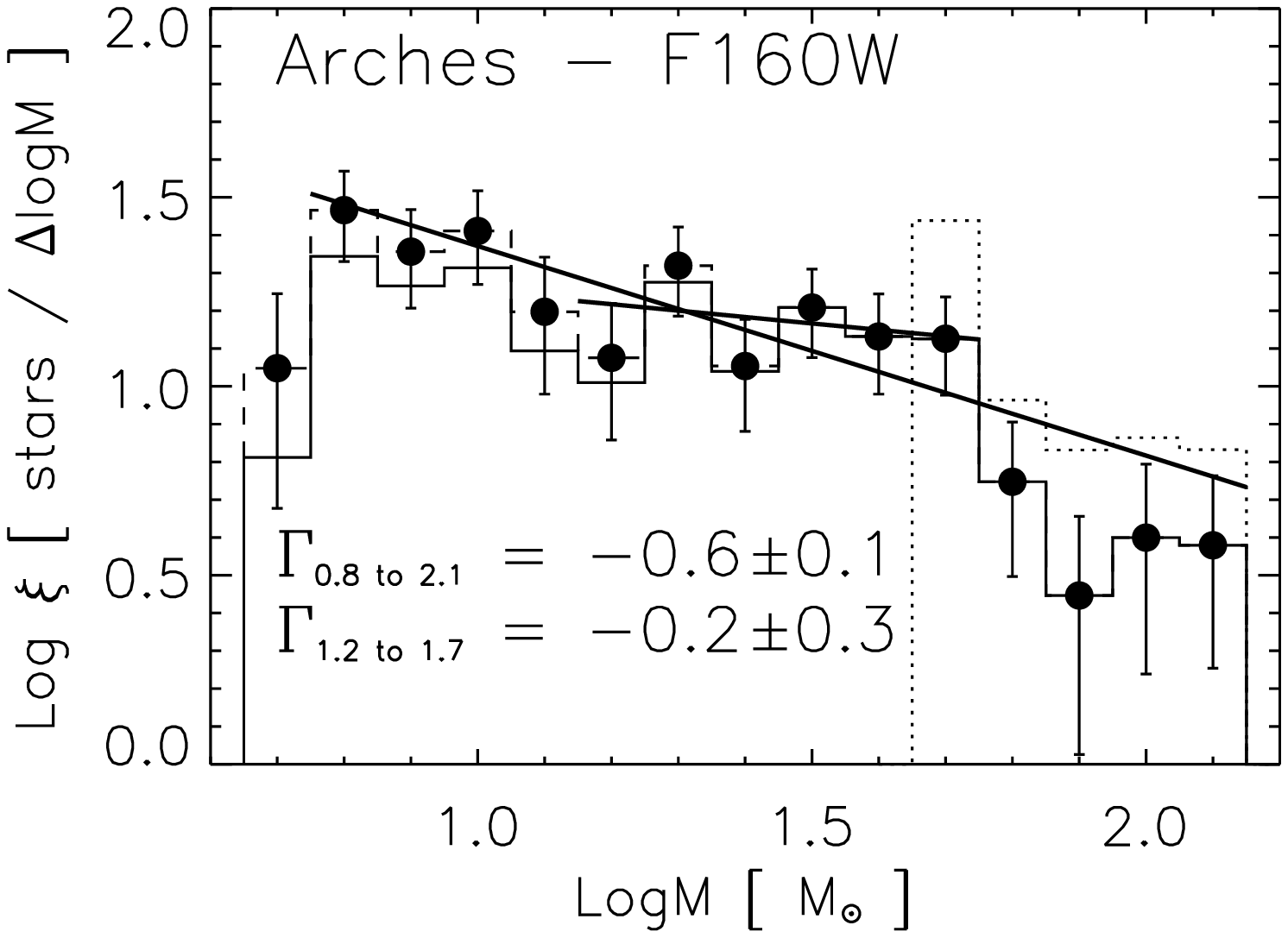}
\end{center}
\caption[]{Mass function for Arches cluster as measured in the F160W image, reproduced from 
Figer et al.\ 1999a. Lines have been fit
over two mass ranges. The slopes of both are relatively flat. The dotted histogram shows the number
of massive stars over the whole cluster, including the inner region.}
\label{eps1}
\end{figure}

There are five possible explanations for the flat slope and apparent plateaus: 
1) inaccuracies in the \Minit\ vs. magnitude relation,
2) inaccuracy in estimating cluster age, metallicity, and/or mass-loss rates, 
3) effects of binaries, 
4) effects of dynamical evolution, and 
5) a real and interesting property of the IMF. 
If any of the first four possibilities is important enough
to affect the slope of the observed mass function, then Figure 1 does not truly 
represent the IMF of the Arches cluster. 

The first three possiblities are unlikely to have an effect and have been addressed by
\cite{fig99a}. 
As for the fourth possibility, mass segregation takes place on a short time scale.
In view of the theoretical results in \cite{kim99a} and \cite{kim00}, \cite{fig99a} divided 
the cluster into two annuli, an inner annulus
(2\farcs5 to 4\farcs5), and an outer annulus (4\farcs5 to 7\farcs5), in order to assess a possible change in
${\rm \Gamma}$ as a function of radius (see Figure 8 in \cite{fig99a}). 
Indeed, there is a significant change of slope from $-$0.2 to $-$0.8 
over this annular range, consistent with 
expectations. Note that these figures show that the
primary result of a flat IMF in the Arches cluster is not sensitive to the choice of annulus.
The fifth possibility is most interesting for its implications 
concerning star formation in the GC.
The unusually flat slope is consistent with the expectations that environmental conditions
near the GC tend to favor the formation of high mass 
stars, relative to star formation taking place
elsewhere in the Galaxy. However, before such a conclusion can be drawn and made quantitative,
dynamical evolutionary effects discussed above need to be carefully modelled. 
As discussed in \cite{mor93}, the characteristics of the interstellar
medium in the GC -- particularly the large turbulent velocities, high 
cloud temperatures, strong magnetic fields, and large tidal forces -- lead 
to a relatively large Jeans mass, compared to that elsewhere in the 
Galaxy. 

\section{Super Star Clusters in the GC}
\cite{fig99a} find a total cluster mass for the Arches cluster 
of $\approx$10$^4$~\Msun\ and about twice this value for the Quintuplet cluster. 
Both clusters are potential 
super star clusters, objects noted for their relative youth ($\tau_{\rm age}$~$<$~20~Myr),
compactness (r$_{\rm half-light}$~$<$~1~pc), and large mass (10$^4$~\Msun\ to 
10$^6$~\Msun) \cite{hof96}. 
The total mass estimate for the Arches cluster
is greater than that of NGC 3603, but it is still below the masses of Globular clusters. 
Using the new estimates, and Table 5 in \cite{fig99b}, it seems that R136,
the Quintuplet cluster and the Arches cluster are similar in mass, 
but the Arches cluster is about an order of magnitude
more dense than R136, $\rho_{\rm Arches}$ $\simgr$ 3(10$^5$)~\Msun~pc$^{-3}$. 
In fact, the core of the Arches cluster appears to be more dense than most globular clusters. 
While these clusters appear to be as massive as small globulars, they will not survive
long enough to become globulars. Indeed, there are no similar, but older, clusters 
that have been identified within the central 50 pc \cite{fig95a}.
Fokker-Planck simulations by \cite{kim99a} 
suggest that these clusters will evaporate in $\simls 10$~Myr due mainly to
strong tidal forces. A result confirmed by \cite{kim99b} and \cite{kim00}. 
In fact, we may be seeing two
snapshots in an otherwise identical evolutionary sequence, i.e., it is possible that the
Quintuplet and Arches clusters are similar in their initial properties 
while the former is currently more extended due to its older age and the effects of dynamical evolution.

\section{Conclusions}
The first direct determination of the IMF has been made in the Galactic Center for
the Arches cluster, and it appears to be significantly flatter than the Salpeter value. 
This result implies that the IMF law is not universal and may be affected by environmental
conditions, although a similar result was found for NGC3603 \cite{eis98}.
We are continuing attempts to minimize possible errors in this determination. For
instance, Keck spectroscopy has recently been obtained to verify the mass-magnitude 
relations. Finally, higher resolution imaging is needed to extend the IMF down to lower 
masses and to determine the lower-mass cutoffs.

\end{document}